\newcounter{myctr}
\def\myitem{\refstepcounter{myctr}\bibfont\noindent\ifnum\themyctr>9\else\phantom{0}\fi\hangindent17pt\themyctr.\enskip}

\documentclass{ws-ijqi}

	\usepackage[T1]{fontenc}  

	\usepackage{amsmath}		

	\usepackage{amssymb}

	\usepackage{mathtools}		

	\usepackage{dsfont}      

	\usepackage{fixmath}     

	\usepackage[caption=false]{subfig}

	\usepackage{graphicx}

	\usepackage{ifthen}

	\usepackage[final,colorlinks=true,linkcolor=blue,citecolor=blue]{hyperref}    

  \renewcommand{\d}[1]{\mathrm{d}#1}
  
  \newcommand{\ii}{\mathrm{i}}

  \newcommand{\abs}[1]{\left\lvert{#1}\right\rvert}
  \newcommand{\ee}[1]{\operatorname{e}^{#1}}
  \newcommand*\defeq{\equiv}
  
  \newlength{\bracewidth}
  
  \newcommand*{\per}{\operatorname{per}}

  \newcommand{\ket}[1]{\vert #1 \rangle}
  
  \newcommand{\bigexpval}[1]{\left\langle #1 \right\rangle}

\renewcommand*\d[1]{d #1\,}

\makeatletter
\newcommand\newsubcommand[3]{\newcommand#1{#2\sc@sub{#3}}}
\def\sc@sub#1{\def\sc@thesub{#1}\@ifnextchar_{\sc@mergesubs}{_{\sc@thesub}}}
\def\sc@mergesubs_#1{_{\sc@thesub#1}}

\newcommand\newsupcommand[3]{\newcommand#1{#2\sc@sup{#3}}}
\def\sc@sup#1{\def\sc@thesup{#1}\@ifnextchar^{\sc@mergesups}{^{\sc@thesup}}}
\def\sc@mergesups^#1{^{\sc@thesup#1}}

\makeatother

\renewcommand*\d[1]{d #1\,}

	\newcommand*\Mports{M}
	\newcommand*\Ndets{N}

	\newcommand*\permut{\sigma}
	
	\newcommand*\detd{d}
	
	\newcommand*\sources{s}

	\newcommand*\SymmGroup[1]{\Sigma_{#1}}

\newcommand*\nfock[1]{n_{#1}}
\newcommand*\nfockSet{ \left\{ \nfock{\sources} \right\}}

\newcommand*\fdist{\xi}
\newcommand*\tdist{\chi}

\newcommand*\ProdMatrix[2]{\mathcal{A}^{(\Ndets \times \Ndets)}_{\ifthenelse{\equal{\unexpanded{#2}}{}}{#1}{#1,#2}}}

\newcommand*\OverlapFactor[2]{G_{\ifthenelse{\equal{\unexpanded{#2}}{}}{#1}{#1,#2}}}

\newcommand*\TotalProbInterval[1]{P^{(\Ndets)}_{\nfockSet}\left( \tdettintervalSet \right)}
\newcommand*\TotalProbIntervalSingle[1]{P^{(\Ndets)}\left( 
\tdettintervalSet \right)}

\newcommand*\TotalProbIntervalPol[1]{P^{(\Ndets)}\left( 
\left \{ \tdet{\detd}, \tinterval{\detd}, \phi_d \right\}; \left \{\theta_s \right \}\right)}

\newcommand*\TotalProbIntervalQubit[1]{P^{(\Ndets)}\left( 
\left \{ \tdet{\detd}, \tinterval{\detd}, j_d  \right\}; \left \{\theta_s \right \}\right)}

\newsupcommand{\aoutadjtmp}{\hat{\mathrm{a}}}{\dagger}

\newsupcommand{\ainadjtmp}{\hat{a}}{\dagger}

\newsupcommand{\ainxiadjtmp}{\hat{a}}{\dagger}

	\newcommand*\Eoutchar{\mathrm{E}}
	\newcommand*\Eout[1]{\hat{\Eoutchar}_{#1}(\tout{#1})}
	\newcommand*\Eoutplus[1]{\hat{\Eoutchar}^{(+)}_{#1}(\tout{#1})}
	\newcommand*\Eoutminus[1]{\hat{\Eoutchar}^{(-)}_{#1}(\tout{#1})}
	\newcommand*\Eoutplusminus[1]{\hat{\Eoutchar}^{(\pm)}_{#1}(\tout{#1})}
	
	\newcommand*\Einchar{E}
	\newcommand*\Ein[2]{\hat{\Einchar}_{#1}\ifthenelse{\equal{\unexpanded{#2}}{}}{}{(\tfromU{#1}{#2})}}
	\newcommand*\Einplus[2]{\hat{\Einchar}^{(+)}_{#1}\ifthenelse{\equal{\unexpanded{#2}}{}}{}{(\tout{#2})}}
	\newcommand*\Einminus[2]{\hat{\Einchar}^{(-)}_{#1}\ifthenelse{\equal{\unexpanded{#2}}{}}{}{(\tout{#2})}}
	\newcommand*\Einplusminus[2]{\hat{\Einchar}^{(\pm)}_{#1}\ifthenelse{\equal{\unexpanded{#2}}{}}{}{(\tout{#2})}}

\newcommand*{\Umatrix}[1]{\mathcal{U}^{(\Ndets \times \Ndets)}_{#1}}
\newcommand*{\UmatrixTot}{\mathcal{U}^{(\Mports \times \Mports)}}

\newcommand*\Uto[2]{\mathcal{U}_{#2,#1}}

\newcommand*\Dto[2]{\mathcal{U}_{#2,#1}}

\newcommand*\CorrMatrix[1]{\mathcal{T}^{(\Ndets \times \Ndets)}_{#1}}

\newcommand\tout[1]{t_{#1}}

\newcommand*\tdet[1]{t_{#1}}

\newcommand*\toutSet{ \left\{ \tout{\detd} \right\}}

\newcommand*\tinterval[1]{\Delta \tdet{#1}}
\newcommand*\tdettintervalSet{ \left\{ \tdet{\detd}, \tinterval{\detd} \right\}}

	\newcommand*\Gn[1]{G^{(N)}_{#1}(\toutSet)}
	\newcommand*\GnP[1]{G^{(N)}_{#1}(\{t_d,\phi_d\};\{\theta_s\})}

\begin{document}

\markboth{Vincenzo Tamma}
{Sampling of bosonic qubits}

\catchline{}{}{}{}{}

\title{Sampling of bosonic qubits}

\author{Vincenzo Tamma}

\address{Institut für Quantenphysik and Center for Integrated Quantum Science and Technology ($IQ^{ST}$),\\ Universität Ulm, D-89069 Ulm, Germany\\
vincenzo.tamma@uni-ulm.de}

\maketitle

\begin{history}
\received{Day Month Year}
\revised{Day Month Year}
\end{history}

\begin{abstract}
The \textit{boson sampling problem}  has brought a lot of attention in the quantum information field because it is not efficiently solvable with a classical computer; nonetheless it can be implemented with linear optical interferometers with single-boson sources. Recently, we introduced  a more general problem, the \textit{multi-boson correlation sampling problem}, which allows us to take advantage of the multi-mode spectral distribution of the bosonic sources together with time-correlated measurements in order to achieve sampling not only over the output ports of the interferometer but also over the joint detection times. This problem was analyzed for both single-photon sources and thermal sources. In this work, we demonstrate that it is possible to use single qubit bosonic sources in order to sample not only over the described ``space-time'' degree of freedom but also over all the possible exponential number of multi-qubit output states.
\end{abstract}


\section{Introduction}
The \emph{boson sampling problem} (BSP) \cite{aaronson2011computational,Franson2013,Ralph2013,Gard2014BS} has recently raised a lot of attention in the research community leading to several proof-of-principle demonstrations  \cite{Broome2013,Crespi2013,tillmann2013experimental,Spring2013},  as well as studies of its characterization \cite{Spagnolo2014,Carolan2013,Tichy2014,Shchesnovich2014,Gard2014,Tillmann2014,deGuise2014,Tan2013} and of its implementation with different types of sources \cite{Rahimi2014,Lund2014,Seshadreesan2014,Motes2014,Olson2014,Rohde2013,Motes2013}.  The BSP problem consists of sampling the probability distribution associated with $\Ndets$ bosons at the output of a linear $\Mports$-port interferometer, where the condition $\Mports \gg \Ndets$ ensures that the probability for the bunching of bosons is reasonably small \cite{aaronson2011computational}. 
The BSP is hard to solve with a classical computer since it involves an exponential number of possible samples, each connected with the permanent of a random complex matrix. Indeed, the calculation of these permanents is a $\# P$-problem \cite{Valiant1979} and therefore not efficiently computable with a classical computer.

In recent works \cite{FockMBCSP,ThermalMBCSP,Tamma2013,Naegele}, we discussed the use of independent sources with arbitrary spectral distributions.  Moreover, we analyzed the joint probability rate associated with the correlated detections of $\Ndets$ bosons in a given sample of $\Ndets$ output ports. Here, differently from the original formulation of the BSP, the sampling problem relies on the times at which the bosons are detected. This leads to the more general \emph{multi-boson correlation sampling problem} (MBCSP), which we analyzed in the case of Fock state sources \cite{FockMBCSP} and thermal sources \cite{ThermalMBCSP}. 

In this paper, we consider the case of arbitrarily polarized single-photon sources and we analyze the sampling problem over all the possible correlated single-photon detections both in time and polarization. By performing measurements in a given polarization base, it will be possible to sample over all the possible output $N$-qubit states (corresponding to $N$-binary-digit numbers) within each sample of $N$ output ports.

The paper is organized as follows: we give a brief summary of the MBCSP for single-photon sources in section~\ref{sec:MBCSP}; we consider the MBCSP based on polarization correlation measurements in section~\ref{sec:PolarizSources}: after analyzing arbitrary polarization correlation measurements in section~\ref{sec:PolarizCorr}, we address the problem of the qubit sampling in section~\ref{sec:QubitSampl};  we conclude with final remarks in section~\ref{sec:Conclusion}.

\section{The multi-boson correlation sampling problem (MBCSP) with multi-mode single-photon sources}\label{sec:MBCSP}
We summarize the MBCSP of a given order $N$ with multi-mode single-photon sources as follows. First, we prepare a random linear interferometer (see Fig.~1) with $M \gg N$ ports. The interferometer input state $\ket{\Psi_N}$ is the $N$-photon state consisting of multi-mode single-photon states \cite{Loudon} 
\begin{align}
	\ket{1}_{\xi_s} = \int_{-\infty}^{+\infty} \d{\omega} \xi_s(\omega) \ainadjtmp_{s}(\omega) \ket{0},
	\label{eqn:SinglePhotonState}
\end{align}
in each of $N$ out of $M$ given input ports $s =1,...,N$, where $\xi(\omega)$ can be for simplicity a generic Gaussian narrow frequency distribution,  and the vacuum in the remaining ports. Secondly, using time-resolving detectors, we sample from all possible correlated detection events in which $\Ndets$ single photons are detected at only $\Ndets$ of the $\Mports$ output ports at any possible sequence $\left\{ \tdet{\detd}\right\}_{d=1}^N $ of correlated detection times    for given integration intervals  $\left\{ \tinterval{\detd}\right\}$. The total sampling time domain must be sufficiently large with respect to the coherence times of each single photon as well as the relative time delays of the photon pulses at the detectors. 

\begin{figure}
	\begin{center}
		\includegraphics{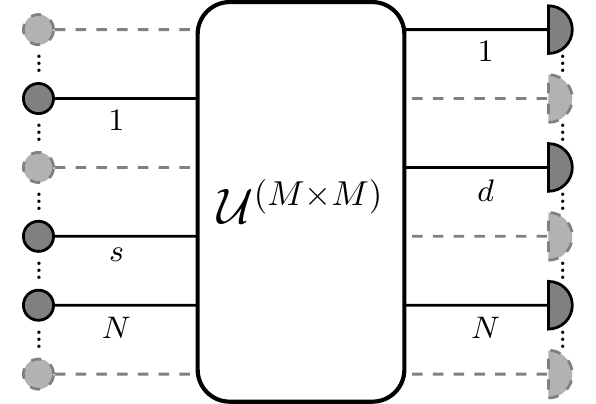}
		\caption{Implementation of the MBCSP of order $N$ with a random linear interferometer with $M \gg N$ ports and  $N$ single-photon sources. After the evolution in the interferometer, described by a unitary random matrix $\UmatrixTot$, correlated detection events are recorded in $\Ndets$ of the $M$ output ports, labeled as $\detd=1,\dots,\Ndets$.}
		\label{fig:LinearInterferometerSingleFock}
	\end{center}
\end{figure}

The rate of $\Ndets$-fold joint detection events for a given sample $\detd=1,\dots,\Ndets$ of output ports is proportional to the $\Ndets$th-order Glauber-Sudarshan correlation function \cite{Glauber2007,Scully1997,Shih2011}
\begin{align}
	\Gn{}  
	=\bigexpval{\Eoutminus{1}\dots\Eoutminus{\Ndets}\Eoutplus{\Ndets}\dots\Eoutplus{1}}_{\ket{\Psi_N}}, 
	\label{eqn:GeneralSetup:CorrelationFunction}
\end{align}
where $\Eoutplusminus{\detd}$ denotes the positive/negative frequency parts of the field operator $\Eout{\detd}=\Eoutplus{\detd} + \Eoutminus{\detd}$ at the $\detd$th detector. These field operators are connected with the field operators at the input ports by a unitary operator $\hat{U}$ that describes the evolution in the linear interferometer. Equivalently, the interferometer can be described by a random unitary $\Mports \times \Mports$ matrix $\UmatrixTot$, which we assume for simplicity to be frequency independent (any unitary matrix $\UmatrixTot$ can be implemented by using only passive linear optical elements \cite{Reck1994}). For a specific set of $N$ occupied input ports and $\Ndets$ output ports where a joint detection occurs, the $\Ndets \times \Ndets$ submatrix 
\begin{align}
	\Umatrix\, \defeq \Big[ \Uto{\sources}{\detd} \Big]_{\substack{\detd=1,\dots,\Ndets \\ \sources=1,\dots,\Ndets}}.
	\label{eqn:SingleFock:Umatrix}
\end{align}
allows us to express the electric field operators at the detectors as linear combinations 
\begin{align}
	\Eoutplus{\detd} = \sum_{\sources=1}^{\Mports} \Uto{\sources}{\detd}\Einplus{\sources}{\detd}
	\label{eqn:GeneralSetup:ExpansionEPlus}
\end{align}
of the field operators $\Einplus{\sources}{\detd}$ at the sources.

In Ref. \cite{FockMBCSP} we found  that the $\Ndets$th-order Glauber-Sudarshan correlation function can be expressed in terms of matrices
\begin{align}
				\CorrMatrix{} \defeq 
		\big[ \Dto{\sources}{\detd} \ \tdist_{\sources}(\tout{\detd})
		\big]_{\substack{\detd=1,\dots,\Ndets \\ \sources=1,\dots,\Ndets}} ,
		\label{eqn:CorrMatrixDefintion}
\end{align}
given by the Hadamard product of the interferometer matrix $\Umatrix\,$ in Eq. (\ref{eqn:SingleFock:Umatrix}) and the time-dependent matrix $[\tdist_{\sources}(t_d)]_{\substack{\detd=1,\dots,\Ndets \\ \sources=1,\dots,\Ndets}}$, defined by the Fourier transforms
\begin{align}
	\tdist_{\sources}(t) \defeq \frac{1}{\sqrt{2\pi}} \int_{-\infty}^{+\infty} \fdist_{\sources}
	(\omega) \ee{-\ii \omega t} \d{\omega}
	\label{eqn:TemporalDistribution} 
\end{align}
of the single-photon distributions $\xi_s(\omega)$, with $s = 1,...,N$, at times $t=t_d$, with $d=1,...,N$ .
Indeed, in the  narrow-bandwidth approximation Eq. (\ref{eqn:GeneralSetup:CorrelationFunction}) becomes \cite{FockMBCSP} 
\begin{align}
	\Gn{} 
	&\propto \abs{ \sum_{\permut\in \SymmGroup{\Ndets}}^{}
	\prod_{\sources=1}^{\Ndets}\, \big[\Uto{\sources}{\permut(\sources)} \,
	\tdist_{\sources}(\tout{\permut(\sources)})\big] }^2
	\label{eqn:CorrelationFinal2}
	\\
	&=  \abs{\per \CorrMatrix{}}^2 
	.
	\label{eqn:CorrelationFinal}
\end{align}
Here, the permanent $\per \CorrMatrix{}$ is defined as the sum over all the permutations $\sigma$ in the symmetric group $\Sigma_N$ corresponding to all $N!$ amplitudes $\prod_{\sources=1}^{\Ndets}  [\Uto{\sources}{\permut(\sources)}\,
	\tdist_{\sources}(\tout{\permut(\sources)})]$ associated with each possible multi-photon quantum path from the $N$ sources to the $N$ detectors.

In the case of ideal detectors we easily obtain, by using Eq. (\ref{eqn:CorrelationFinal}), the probability 
\begin{align}
	\TotalProbIntervalSingle{} 
=\int_{\tdet{1} -
	\tinterval{1}/2}^{\tdet{1}+\tinterval{1}/2} \d{\tdet{1}} \dots
	\int_{\tdet{\Ndets}-\tinterval{\Ndets}/2}^{\tdet{\Ndets}+\tinterval{\Ndets}/2}\d{\tdet{\Ndets}}
	\abs{\per \CorrMatrix{}}^2 	
	\label{eqn:Probability}
\end{align}
for an $N$-fold joint detection event in the output ports $d=1,2,...,N$ and in a given sample of time intervals $\left\{ \left[
\tdet{\detd} - \tinterval{\detd}/2, \tdet{\detd} + \tinterval{\detd}/2 \right]
\right\}_{d=1,2,...,N}$.
%
%
%
%
%

\section{Boson Sampling based on polarization correlation measurements}\label{sec:PolarizSources}

In this section, we show how we can take advantage of the polarization degree of freedom in order to implement a sampling problem not only in the output spatial channels and the detection times but also in the measured correlated polarizations of the output photons.
\begin{figure}
	\begin{center}
		\includegraphics{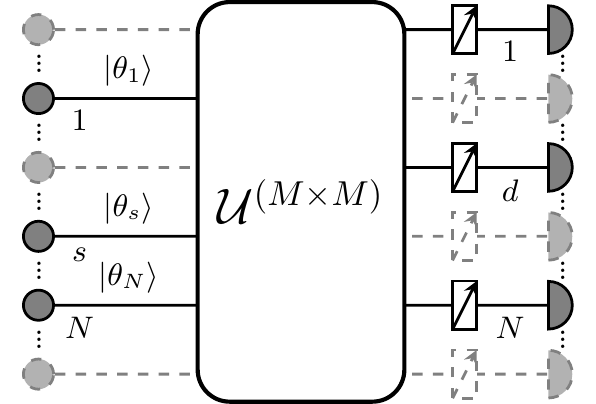}
		\caption{Implementation of the bosonic qubit sampling problem with a random linear interferometer with $M \gg N$ ports and  $N$ single-qubit sources with polarization states $\vert\theta_s \rangle $, where $s = 1,2,...,N$. After the evolution in the interferometer, described by a unitary random matrix $\UmatrixTot$, correlated detection events both in time and polarization are recorded in $N$ of the $M$ output ports, labeled as $\detd=1,\dots,\Ndets$. The polarization correlation measurements are performed by using polarization analyzers with generic orientations in front of the single photon detectors. Moreover, for a defined polarization base $\{\vert H \rangle, \vert V \rangle\}$, it is possible for each port to measure a qubit either in the state $\vert H \rangle $ or $\vert V \rangle $ by using a polarization beam splitter and two detectors recording either the transmitted H-polarized qubit or the reflected V-polarized qubit.}
		\label{fig:LinearInterferometerSingleFock}
	\end{center}
\end{figure}

For this reason we consider (see Fig. 2) $N$ multi-mode single photon sources in the input ports $s=1,2,..N$, with polarizations $\theta_s$, corresponding, in a given  base $\{\vert H \rangle, \vert V \rangle\} \equiv \{\vert 0 \rangle, \vert 1 \rangle\}$, to $N$ generic polarization qubits 
\begin{align}
\vert \theta_s \rangle = \cos \theta_s \vert 0 \rangle + \sin \theta_s \vert 1 \rangle,
\label{QubitStates}
\end{align}
leading to the $N$-qubits state
\begin{eqnarray}
\vert \Psi_{in} \rangle \equiv \bigotimes_{s=1}^{N} \vert \theta_{s} \rangle = \sum_{i_1,...,i_s,...,i_N=0,1}\bigg[\prod_{s=1}^N \cos (\theta_s - \frac{\pi}{2} i_s)\bigg]  \vert i_1...i_s...i_N \rangle,
\label{NQubitState}
\end{eqnarray}
with $\vert i_1...i_s...i_N \rangle \equiv \bigotimes_{s=1}^{N} \vert i_s \rangle$. This state is given by the superposition of an exponential number $2^N$ of $N$-qubit states $\vert i \rangle \equiv \vert i_1...i_s...i_N \rangle$, with $i = 0,1,..,2^N - 1$ with amplitudes determined by the generic polarization angles $\theta_s$, with $ s=1,2,...N$.

After the evolution of these qubits in the interferometer, that we assume for simplicity to be polarization independent, polarization correlation measurements are performed by sampling over all $M$ output ports.

\subsection{Correlation measurements at arbitrary polarization angles}\label{sec:PolarizCorr}

We consider first the case of correlation measurements at given angles $\phi_d$ of the polarization analyzers at each of the output ports, $d=1,2,...,N$, where a joint-detection occurs. Here, it is useful to introduce the ``effective'' matrix
\begin{align}
\mathcal{B}^{(\phi_d,\theta_s)}  \equiv \Big[ \cos (\phi_d - \theta_s) \,  \Uto{\sources}{\detd} \Big]_{\substack{\detd=1,\dots,\Ndets \\ \sources=1,\dots,\Ndets}}
\label{Bmatrix}
\end{align}
obtained by the Hadamard product of the matrix $\Umatrix \,$ in Eq. (\ref{eqn:SingleFock:Umatrix})
and the polarization matrix $[\cos (\phi_d - \theta_s)]_{\substack{\detd=1,\dots,\Ndets \\ \sources=1,\dots,\Ndets}}$, defined by the source polarizations $\{\theta_s\}$ and the detected polarizations $\{\phi_d\}$.

By substituting the polarization-independent interferometric matrix $\Umatrix\, $ in Eq. (\ref{eqn:SingleFock:Umatrix}) with the ``effective'' matrix (\ref{Bmatrix}) the $N$-order correlation function in Eq. (\ref{eqn:CorrelationFinal2}) becomes
\begin{align}
	\GnP{}  
	\propto \abs{\per \,\big[\mathcal{B}^{(\phi_d,\theta_s)}_{d,s}  \chi_s(t_d)\big]_{\substack{\detd=1,\dots,\Ndets \\ \sources=1,\dots,\Ndets}}}^2.
	\label{eqn:GeneralSetup:CorrelationFunction2}
\end{align}

In an analogous way,  Eq. (\ref{eqn:Probability}) corresponds to 
\begin{align}
	&\TotalProbIntervalPol{} = \nonumber \\
	 & \int_{\tdet{1} -
	\tinterval{1}/2}^{\tdet{1}+\tinterval{1}/2} \d{\tdet{1}} \dots
	\int_{\tdet{\Ndets}-\tinterval{\Ndets}/2}^{\tdet{\Ndets}+\tinterval{\Ndets}/2}\d{\tdet{\Ndets}}
	\abs{\per \,\big[\mathcal{B}^{(\phi_d,\theta_s)}_{d,s}  \chi_s(t_d)\big]_{\substack{\detd=1,\dots,\Ndets \\ \sources=1,\dots,\Ndets}}}^2,	
	\label{eqn:Probability2}
\end{align}
representing the $N$-fold joint detection probability of measuring the $N$ qubits with polarization angles $\phi_d$, for a given sample $d=1,2,...N$ of output modes and detection time intervals  $\left\{ \left[
\tdet{\detd} - \tinterval{\detd}/2, \tdet{\detd} + \tinterval{\detd}/2 \right]
\right\}$.

%
%

In conclusion, we have obtained a compact expression in terms of permanents for the probability of detecting the $N$ qubits in each possible sample of output ports, depending not only on the detector integration intervals but also on the general orientation of the polarization analyzers in front of the detectors.

\subsection{Sampling of photonic qubits}\label{sec:QubitSampl}

In this section, we finally describe how the quantum network in Fig. 2 can be used to implement a multi-boson correlation sampling problem, where the sampling occurs not only over the output spatial channels and the joint-detection times but also over all possible output $N$-photon polarization outcomes, with respect to the  base $\{\vert H \rangle, \vert V \rangle\} \equiv \{\vert 0 \rangle, \vert 1 \rangle\}.$  In order to perform correlated polarization measurements in such a base we use, for each output port, a polarization beam splitter and two detectors recording either a transmitted $H$-polarized qubit or a reflected $V$-polarized qubit.  Thereby, it is possible for each port to measure a qubit either in the state $\vert 0 \rangle $ or $\vert 1 \rangle$. Indeed, our device is sampling over all the possible qubit correlation measurements with outcomes $\{\vert j_d \rangle\}_{d=1,...,N}$, where  $j_d = 0,1$.

Depending on the measured qubit states $\{\vert j_d \rangle\}_{d=1,...,N}$ for any possible ``spatial'' sample $d=1,2,...,N$ the ``effective'' matrices (\ref{Bmatrix}) reduce to 
\begin{align}
\mathcal{B}^{(j_d,\theta_s)}  \equiv \Big[ \cos ( \frac{\pi}{2} j_d - \theta_s) \,  \Uto{\sources}{\detd} \Big]_{\substack{\detd=1,\dots,\Ndets \\ \sources=1,\dots,\Ndets}},
\label{Bmatrix2}
\end{align}
and the polarization correlation probability in Eq. (\ref{eqn:Probability2}) becomes
\begin{align}
	&\TotalProbIntervalQubit{} = \nonumber \\
	 & \int_{\tdet{1} -
	\tinterval{1}/2}^{\tdet{1}+\tinterval{1}/2} \d{\tdet{1}} \dots
	\int_{\tdet{\Ndets}-\tinterval{\Ndets}/2}^{\tdet{\Ndets}+\tinterval{\Ndets}/2}\d{\tdet{\Ndets}}
	\abs{\per \,\big[\mathcal{B}^{(j_d,\theta_s)}_{d,s}  \chi_s(t_d)\big]_{\substack{\detd=1,\dots,\Ndets \\ \sources=1,\dots,\Ndets}}}^2,	
	\label{eqn:Probability3}
\end{align} 
for the $N$-qubit input state in Eq. (\ref{NQubitState}). 
Here, the $N!$ amplitudes in the permanent in Eq. (\ref{eqn:Probability3}), associated with all the interfering multi-photon quantum paths, depend not only on the interferometer evolution, the source frequency distribution and detection time intervals, but also on which of the possible $2^N$ qubit states  $\vert j_1 ...j_d ...j_N \rangle$ is detected in the qubit sampling problem.

\section{Conclusion}\label{sec:Conclusion}

In this paper, we generalized the multi-boson correlation sampling problem to photonic qubits, by taking advantage of the polarization degree of freedom of the input photons.  Given a linear interferometer with single-photon qubit sources, we defined a more general problem based on the sampling of the output probability distribution not only in the  spatial channels and associated detection intervals but also in all possible qubit measurement outcomes, exponential in number, in a given polarization base. We found, for each possible degree of freedom in the sampling, i.e. spatial channels, time, and polarization, a corresponding matrix. It is the modulo squared permanent of the Hadamard product of these matrices which determines the detection probability rate for each possible sample. Finally, the probability associated with each correlation measurement emerges naturally from integrating over the  detection time intervals associated with the time sampling. 

The result obtained so far can be easily extended to thermal sources \cite{ThermalMBCSP} and, in general, from photonic networks to atomic interferometers with bosonic qubits, where, for example, the spin degree of freedom takes the place of the polarization.
Although for simplicity we have addressed the case of polarization-independent evolution through the interferometer, our work can be generalized to any non-trivial interferometric evolution in the multi-qubit Hilbert space. 

This research may pave the way towards the developments of quantum networks able to achieve quantum speed-up in useful computational problems beyond boson sampling.

\section{Acknowledgments}
I would like to thank  M. Freyberger, S. Laibacher, F. N\"{a}gele,  W. P. Schleich, and J. Seiler as well as J. Franson, S. Lomonaco, T. Pittmann,  and Y.H. Shih for fruitful discussions during my visit at UMBC in the summer of 2013. 
I also acknowledge the support of the German Space Agency DLR with funds provided by the Federal Ministry of Economics and Technology (BMWi) under grant no. DLR 50 WM 1136.

\end{document}